\documentclass[12pt,a4paper]{article}
\makeatletter
\def\eqnarray{%
\stepcounter{equation}%
\let\@currentlabel=\theequation
\global\@eqnswtrue
\global\@eqcnt\z@
\tabskip\@centering
\let\\=\@eqncr
$$\halign to \displaywidth\bgroup\@eqnsel\hskip\@centering
$\displaystyle\tabskip\z@{##}$&\global\@eqcnt\@ne
\hfil$\displaystyle{{}##{}}$\hfil
&\global\@eqcnt\tw@$\displaystyle\tabskip\z@{##}$\hfil
\tabskip\@centering&\llap{##}\tabskip\z@\cr}
\makeatother
\newcommand{\fukuso}{{\mathbf C}}

\begin{document}

\title{\sl Explicit Form of Solution of Two Atoms Tavis--Cummings Model}
\author{
  Kazuyuki FUJII 
  \thanks{E-mail address : fujii@yokohama-cu.ac.jp },
  Kyoko HIGASHIDA 
  \thanks{E-mail address : s035577d@yokohama-cu.ac.jp },
  Ryosuke KATO 
  \thanks{E-mail address : s035559g@yokohama-cu.ac.jp }, 
  Yukako WADA 
  \thanks{E-mail address : s035588a@yokohama-cu.ac.jp }\\
  Department of Mathematical Sciences\\
  Yokohama City University\\
  Yokohama, 236--0027\\
  Japan
  }
\date{}
\maketitle
%
%
%
%
\begin{abstract}
  In this paper we consider the two atoms Tavis--Cummings model and 
  give an explicit form to the solution of this model which will play a 
  central role in quantum computation based on atoms of laser--cooled and 
  trapped linearly in a cavity. 
  
  We also present a problem of three atoms Tavis--Cummings model which is 
  related to the construction of controlled--controlled NOT operation (gate) 
  in quantum computation.
\end{abstract}
%


%
%
%
%

\newpage

The purpose of this paper is to give an explicit form to the solution of 
Tavis--Cummings model (\cite{TC}) with one and two atoms. This model is a very 
important one in Quantum Optics and has been studied widely , see \cite{AE}, 
\cite{MSIII} or \cite{C-HDG} as general textbooks in quantum optics. 
See also recent papers \cite{GBR}, \cite{MFr} and their references.

We are studying a quantum computation and therefore want to study the model 
from this point of view, namely the quantum computation based on atoms of 
laser--cooled and trapped linearly in a cavity. We must in this model 
construct a controlled NOT gate or other controlled unitary gates to perform 
a quantum computation, see \cite{KF1} as a general introduction to 
this subject. 

For that purpose we need the explicit form of solution of the models with 
one, two and three atoms. As for the model of one atom it is more or less 
well-known, and as for the case of two or three atoms it has not been given 
as far as we know. 

In this paper we give it for the case of two atoms, while we could not give it 
for the three atoms case, so we present it as a challenging problem. 
Anyway, let us start. 

The Tavis--Cummings model (with $n$--atoms) that we will treat in this paper 
can be written as follows (we set $\hbar=1$ for simplicity). 
\begin{equation}
\label{eq:hamiltonian}
H=
\omega {1}_{L}\otimes a^{\dagger}a + 
\frac{\Delta}{2} \sum_{i=1}^{n}\sigma^{(3)}_{i}\otimes {\bf 1} +
g\sum_{i=1}^{n}\left(
\sigma^{(+)}_{i}\otimes a+\sigma^{(-)}_{i}\otimes a^{\dagger} \right),
\end{equation}
where $\omega$ is the frequency of radiation field, $\Delta$ the energy 
difference of two level atoms, $a$ and $a^{\dagger}$ are 
annihilation and creation operators of the field, and $g$ a coupling constant, 
and $L=2^{n}$. Here $\sigma^{(+)}_{i}$, $\sigma^{(-)}_{i}$ and 
$\sigma^{(3)}_{i}$ are given as 
\begin{equation}
\sigma^{(s)}_{i}=
1_{2}\otimes \cdots \otimes 1_{2}\otimes \sigma_{s}\otimes 1_{2}\otimes \cdots 
\otimes 1_{2}\ (i-\mbox{position})\ \in \ M(L,\fukuso)
\end{equation}
where $s$ is $+$, $-$ and $3$ respectively and 
\begin{equation}
\label{eq:sigmas}
\sigma_{+}=
\left(
  \begin{array}{cc}
    0& 1 \\
    0& 0
  \end{array}
\right), \quad 
\sigma_{-}=
\left(
  \begin{array}{cc}
    0& 0 \\
    1& 0
  \end{array}
\right), \quad 
\sigma_{3}=
\left(
  \begin{array}{cc}
    1& 0  \\
    0& -1
  \end{array}
\right), \quad 
1_{2}=
\left(
  \begin{array}{cc}
    1& 0  \\
    0& 1
  \end{array}
\right).
\end{equation}

Here let us rewrite the hamiltonian (\ref{eq:hamiltonian}). If we set 
\begin{equation}
\label{eq:large-s}
S_{+}=\sum_{i=1}^{n}\sigma^{(+)}_{i},\quad 
S_{-}=\sum_{i=1}^{n}\sigma^{(-)}_{i},\quad 
S_{3}=\frac{1}{2}\sum_{i=1}^{n}\sigma^{(3)}_{i},
\end{equation}
then (\ref{eq:hamiltonian}) can be written as 
\begin{equation}
\label{eq:hamiltonian-2}
H=
\omega {1}_{L}\otimes a^{\dagger}a + \Delta S_{3}\otimes {\bf 1} + 
g\left(S_{+}\otimes a + S_{-}\otimes a^{\dagger} \right)
\equiv H_{0}+V,
\end{equation}
which is very clear. We note that $\{S_{+},S_{-},S_{3}\}$ satisfy the 
$su(2)$--relation 
\begin{equation}
[S_{3},S_{+}]=S_{+},\quad [S_{3},S_{-}]=-S_{-},\quad [S_{+},S_{-}]=2S_{3}.
\end{equation}
However, the representation $\rho$ defined by 
\[
\rho(\sigma_{+})=S_{+},\quad \rho(\sigma_{-})=S_{-},\quad 
\rho(\sigma_{3}/2)=S_{3}
\]
is a reducible representation of $su(2)$. 

We would like to solve the Schr{\" o}dinger equation 
\begin{equation}
\label{eq:schrodinger}
i\frac{d}{dt}U=HU=\left(H_{0}+V\right)U, 
\end{equation}
where $U$ is a unitary operator. 
We can solve this equation by using the {\bf method of constant variation}. 
Let us make a brief review. The equation $i\frac{d}{dt}U=H_{0}U$ is easily 
solved to be 
\[
U(t)=\left(\mbox{e}^{-it\Delta S_{3}}\otimes \mbox{e}^{-it\omega N}\right)
U_{0}
\]
where $N=a^{\dagger}a$ is a number operator and $U_{0}$ a constant unitary. 
By changing $U_{0}$ $\longmapsto$ $U_{0}(t)$ and substituting into 
(\ref{eq:schrodinger}) we have the equation 
\begin{equation}
\label{eq:reduction-equation}
i\frac{d}{dt}U_{0}=g\left\{\mbox{e}^{it(\Delta-\omega)}S_{+}\otimes a + 
\mbox{e}^{-it(\Delta-\omega)}S_{-}\otimes a^{\dagger}\right\}U_{0}
\end{equation}
after some algebras. Here let us assume the resonance condition 
\begin{equation}
\label{eq:resonance}
\Delta=\omega,
\end{equation}
which makes the situation simpler. Under this condition the solution of 
(\ref{eq:reduction-equation}) becomes 
\begin{equation}
\label{eq:main-equation}
U_{0}(t)=
\mbox{e}^{-itg\left(S_{+}\otimes a + S_{-}\otimes a^{\dagger}\right)},
\end{equation}
so that the full solution of (\ref{eq:schrodinger}) is given by 
\begin{equation}
\label{eq:full-solution}
U(t)=\left(\mbox{e}^{-it\omega S_{3}}\otimes \mbox{e}^{-it\omega N}\right)
\mbox{e}^{-itg\left(S_{+}\otimes a + S_{-}\otimes a^{\dagger}\right)},
\end{equation}
where we have dropped the constant unitary operator for simplicity. Therefore 
we have only to calculate the term (\ref{eq:main-equation}) explicitly, 
which is however a very hard task \footnote{the situation is very similar to 
that of \cite{FHKW2}}. 
In the following we set 
\begin{equation}
\label{eq:A}
A=S_{+}\otimes a + S_{-}\otimes a^{\dagger}
\end{equation}
for simplicity. 
We can determine\ $\mbox{e}^{-itgA}$\ for $n=1$ (one atom case) and 
$n=2$ (two atoms case) completely. Let us show. 

\vspace{5mm}
\par \noindent 
{\bf One Atom Case}\quad In this case $A$ in (\ref{eq:A}) is written as 
\begin{equation}
\label{eq:A-one}
A=
\left(
  \begin{array}{cc}
    0&           a \\
    a^{\dagger}& 0
  \end{array}
\right).
\end{equation}
Since 
\begin{equation}
\label{eq:relation-one}
A^{2}=
\left(
  \begin{array}{cc}
    aa^{\dagger}&   0          \\
    0           & a^{\dagger}a 
  \end{array}
\right)=
\left(
  \begin{array}{cc}
    N+1& 0  \\
    0  & N
  \end{array}
\right)\equiv D
\end{equation}
with the number operator $N$ and 
\[
A^{2j}=D^{j},\quad A^{2j+1}=D^{j}A,\quad \mbox{for}\quad j\geq 0, 
\]
so we have 
\begin{eqnarray}
\label{eq:solution-one}
\mbox{e}^{-itgA}
&=&
\sum_{j=0}^{\infty}\frac{(-itg)^{2j}}{(2j)!}A^{2j}+
\sum_{j=0}^{\infty}\frac{(-itg)^{2j+1}}{(2j+1)!}A^{2j+1} \nonumber \\
&=&
\sum_{j=0}^{\infty}(-1)^{j}\frac{(tg)^{2j}}{(2j)!}D^{j}-i
\sum_{j=0}^{\infty}(-1)^{j}\frac{(tg)^{2j+1}}{(2j+1)!}D^{j}A  \nonumber \\
&=&
\sum_{j=0}^{\infty}(-1)^{j}\frac{(tg\sqrt{D})^{2j}}{(2j)!}-i\frac{1}{\sqrt{D}}
\sum_{j=0}^{\infty}(-1)^{j}\frac{(tg\sqrt{D})^{2j+1}}{(2j+1)!}A  \nonumber \\
&=&
\left(
  \begin{array}{cc}
  \mbox{cos}\left(tg\sqrt{N+1}\right)& 
  -i\frac{\mbox{sin}\left(tg\sqrt{N+1}\right)}{\sqrt{N+1}}a  \\
  -i\frac{\mbox{sin}\left(tg\sqrt{N}\right)}{\sqrt{N}}a^{\dagger}& 
  \mbox{cos}\left(tg\sqrt{N}\right)
  \end{array}
\right).
\end{eqnarray}
We obtained the explicit form of solution. However, this form is more or less 
well--known, see for example \cite{MSIII}, Chapter 14. 
Because this solution is very convenient, there are many applications, 
see the textbook \cite{MSIII}. 

We note that (\ref{eq:solution-one}) can be decomposed as 
\begin{eqnarray}
&&\left(
  \begin{array}{cc}
  \mbox{cos}\left(tg\sqrt{N+1}\right)& 
  -i\frac{\mbox{sin}\left(tg\sqrt{N+1}\right)}{\sqrt{N+1}}a  \\
  -i\frac{\mbox{sin}\left(tg\sqrt{N}\right)}{\sqrt{N}}a^{\dagger}& 
  \mbox{cos}\left(tg\sqrt{N}\right)
  \end{array}
\right)       
=
\left(
  \begin{array}{cc}
  \mbox{cos}\left(tg\sqrt{N+1}\right)& 
  -i\frac{\mbox{sin}\left(tg\sqrt{N+1}\right)}{\sqrt{N+1}}a  \\
  -ia^{\dagger}\frac{\mbox{sin}\left(tg\sqrt{N+1}\right)}{\sqrt{N+1}}& 
  \mbox{cos}\left(tg\sqrt{N}\right)
  \end{array}
\right)         \nonumber \\
=&&
\left(
  \begin{array}{cc}
  1 & 0 \\
  -ia^{\dagger}\frac{\mbox{tan}\left(tg\sqrt{N+1}\right)}{\sqrt{N+1}} & 1
  \end{array}
\right)
\left(
  \begin{array}{cc}
  \mbox{cos}\left(tg\sqrt{N+1}\right) & 0 \\
  0 & \frac{1}{\mbox{cos}\left(tg\sqrt{N}\right)}
  \end{array}
\right)
\left(
  \begin{array}{cc}
  1 & -i\frac{\mbox{tan}\left(tg\sqrt{N+1}\right)}{\sqrt{N+1}}a \\
  0 & 1
  \end{array}
\right).            \nonumber \\
=&&
\left(
  \begin{array}{cc}
  1 & 0 \\
  -i\frac{\mbox{tan}\left(tg\sqrt{N}\right)}{\sqrt{N}}a^{\dagger} & 1
  \end{array}
\right)
\left(
  \begin{array}{cc}
  \mbox{cos}\left(tg\sqrt{N+1}\right) & 0 \\
  0 & \frac{1}{\mbox{cos}\left(tg\sqrt{N}\right)}
  \end{array}
\right)
\left(
  \begin{array}{cc}
  1 & -i\frac{\mbox{tan}\left(tg\sqrt{N+1}\right)}{\sqrt{N+1}}a \\
  0 & 1
  \end{array}
\right).            \nonumber \\
&&
\end{eqnarray}
We leave the check to the readers. 

This is a Gauss decomposition of unitary operator. This may be used to 
construct a theory of ``quantum" representation of a non--commutative group, 
which is now under consideration.

\vspace{5mm}
\par \noindent 
{\bf Two Atoms Case}\quad In this case $A$ in (\ref{eq:A}) is written as 
\begin{equation}
\label{eq:A-two}
A=
\left(
  \begin{array}{cccc}
    0 &          a & a &           0  \\
    a^{\dagger}& 0 & 0 &           a  \\
    a^{\dagger}& 0 & 0 &           a  \\
    0 & a^{\dagger}& a^{\dagger} & 0
  \end{array}
\right).
\end{equation}
We would like to look for the explicit form of solution like 
(\ref{eq:solution-one}), so we must find a relation like 
(\ref{eq:relation-one}). It is not difficult to see 
\begin{eqnarray}
\label{eq:relation-two-2}
A^{2}&=&
\left(
  \begin{array}{cccc}
    2(N+1) & 0    & 0    & 2a^{2}    \\
    0      & 2N+1 & 2N+1 & 0         \\
    0      & 2N+1 & 2N+1 & 0         \\
    2(a^{\dagger})^{2} & 0 & 0 & 2N 
  \end{array}
\right),                  \\
\label{eq:relation-two-3}
A^{3}&=&2
\left(
  \begin{array}{cccc}
    0   &    (2N+3)a & (2N+3)a &       0           \\
    (2N+1)a^{\dagger}& 0 & 0 &    (2N+1)a          \\
    (2N+1)a^{\dagger}& 0 & 0 &    (2N+1)a          \\
    0   & (2N-1)a^{\dagger}& (2N-1)a^{\dagger} & 0
  \end{array}
\right).
\end{eqnarray}
From (\ref{eq:relation-two-3}) we find a clear relation 
\begin{equation}
\label{eq:clear-relation-two}
A^{3}=
\left(
  \begin{array}{cccc}
    2(2N+3)&        &        &         \\
           & 2(2N+1)&        &         \\
           &        & 2(2N+1)&         \\
           &        &        & 2(2N-1)
  \end{array}
\right)
\left(
  \begin{array}{cccc}
    0 &          a & a &           0  \\
    a^{\dagger}& 0 & 0 &           a  \\
    a^{\dagger}& 0 & 0 &           a  \\
    0 & a^{\dagger}& a^{\dagger} & 0
  \end{array}
\right)\equiv DA.
\end{equation}
This is our key observation. From this it is easy to see 
\[
A^{2j}=D^{j-1}A^{2}\quad \mbox{for}\quad j \geq 1,\quad 
A^{2j+1}=D^{j}A\quad \mbox{for}\quad j \geq 0,
\]
so that we have 
\begin{eqnarray}
\label{eq:solution-two}
\mbox{e}^{-itgA}
&=&{\bf 1}+
\sum_{j=1}^{\infty}\frac{(-itg)^{2j}}{(2j)!}A^{2j}+
\sum_{j=0}^{\infty}\frac{(-itg)^{2j+1}}{(2j+1)!}A^{2j+1} \nonumber \\
&=&{\bf 1}+
\sum_{j=1}^{\infty}(-1)^{j}\frac{(tg)^{2j}}{(2j)!}D^{j-1}A^{2}-i
\sum_{j=0}^{\infty}(-1)^{j}\frac{(tg)^{2j+1}}{(2j+1)!}D^{j}A   \nonumber \\
&=&{\bf 1}+
\frac{1}{D}\sum_{j=1}^{\infty}(-1)^{j}\frac{(tg)^{2j}}{(2j)!}D^{j}A^{2}-i
\sum_{j=0}^{\infty}(-1)^{j}\frac{(tg)^{2j+1}}{(2j+1)!}D^{j}A   \nonumber \\
&=&{\bf 1}+
\frac{1}{D}\sum_{j=1}^{\infty}(-1)^{j}\frac{(tg\sqrt{D})^{2j}}{(2j)!}A^{2}
-i\frac{1}{\sqrt{D}}
\sum_{j=0}^{\infty}(-1)^{j}\frac{(tg\sqrt{D})^{2j+1}}{(2j+1)!}A  \nonumber \\
&=&{\bf 1}+
\frac{1}{D}\left\{-{\bf 1}+\mbox{cos}\left(tg\sqrt{D}\right)\right\}A^{2}
-i\frac{1}{\sqrt{D}}\mbox{sin}\left(tg\sqrt{D}\right)A
\end{eqnarray}
or more explicitly 
\begin{equation}
\label{eq:solution-two-more}
\mbox{e}^{-itgA}=
\left(
  \begin{array}{cccc}
    a_{11} & a_{12} & a_{13} & a_{14} \\
    a_{21} & a_{22} & a_{23} & a_{24} \\
    a_{31} & a_{32} & a_{33} & a_{34} \\
    a_{41} & a_{42} & a_{43} & a_{44} 
  \end{array}
\right)
\end{equation}
where 
\begin{eqnarray}
a_{11}&=&\frac{N+2+(N+1)\mbox{cos}\left(tg\sqrt{2(2N+3)}\right)}{2N+3},\quad 
a_{12}=a_{13}=
-i\frac{\mbox{sin}\left(tg\sqrt{2(2N+3)}\right)}{\sqrt{2(2N+3)}}a, \nonumber \\
a_{14}&=&\frac{-1+\mbox{cos}\left(tg\sqrt{2(2N+3)}\right)}{2N+3}a^{2},\quad 
a_{21}=a_{31}=
-i\frac{\mbox{sin}\left(tg\sqrt{2(2N+1)}\right)}{\sqrt{2(2N+1)}}a^{\dagger}, 
\nonumber \\
a_{22}&=&a_{33}=\frac{1+\mbox{cos}\left(tg\sqrt{2(2N+1)}\right)}{2},\quad 
a_{23}=a_{32}=\frac{-1+\mbox{cos}\left(tg\sqrt{2(2N+1)}\right)}{2}, 
\nonumber \\
a_{41}&=&
\frac{-1+\mbox{cos}\left(tg\sqrt{2(2N-1)}\right)}{2N-1}{(a^{\dagger})^2},\quad 
a_{42}=a_{43}=
-i\frac{\mbox{sin}\left(tg\sqrt{2(2N-1)}\right)}{\sqrt{2(2N-1)}}a^{\dagger}, 
\nonumber \\
a_{44}&=&\frac{N-1+N\mbox{cos}\left(tg\sqrt{2(2N-1)}\right)}{2N-1}. \nonumber
\end{eqnarray}
This is our main result in this paper. The explicit form has not been known 
in the literature as far as we know. 

Since the Tavis--Cummings model has a kind of universal characteristic and 
the explicit form of solution was given, there must be many applications to 
Quantum Optics, Mathematical Physics and etc. In the forthcoming paper 
\cite{FHKW3} we will apply this to the construction of controlled--unitary 
operations (gates) in quantum computation (see for example \cite{KF1}) based 
on atoms of laser--cooled and trapped linearly in a cavity.

\vspace{5mm}
\par \noindent 
{\bf Three Atoms Case}\quad In this case $A$ in (\ref{eq:A}) is written as 
\begin{equation}
\label{eq:A-three}
A=
\left(
  \begin{array}{cccccccc}
    0 &          a & a &           0  & a & 0 & 0 & 0          \\
    a^{\dagger}& 0 & 0 &           a  & 0 & a & 0 & 0          \\
    a^{\dagger}& 0 & 0 &           a  & 0 & 0 & a & 0          \\
    0 & a^{\dagger}& a^{\dagger} & 0  & 0 & 0 & 0 & a          \\
    a^{\dagger}& 0 & 0  &  0          & 0 & a & a & 0          \\
    0 & a^{\dagger}& 0  & 0   & a^{\dagger} &  0 & 0 & a       \\
    0 & 0 & a^{\dagger} & 0  & a^{\dagger} &  0 & 0 & a        \\
    0 & 0 & 0 & a^{\dagger} & 0 & a^{\dagger} & a^{\dagger} & 0    
  \end{array}
\right)
=
\left(
  \begin{array}{cc}
    \tilde{A} & a1_{4}            \\
    a^{\dagger}1_{4} & \tilde{A}
  \end{array}
\right),
\end{equation}
where $\tilde{A}$ is $A$ in (\ref{eq:A-two}). 

We would like to look for the explicit form of solution like 
(\ref{eq:solution-one}) or (\ref{eq:solution-two-more}). 
However, we could not find a relation like (\ref{eq:relation-one}) or 
(\ref{eq:clear-relation-two}) for (\ref{eq:A-three}) in spite of much 
effort (we have calculated $A^{2}$, $\cdots$, $A^{5}$). 
We encourage the readers to tackle this problem. 

We note that the solution is deeply related to the construction of 
controlled--controlled NOT operation (gate) in quantum computation, 
so the explicit form of it is needed. 

\vspace{5mm}
We conclude this paper by making a comment on our target. The Tavis--Cummings 
model is based on two energy levels of atoms. However, an atom has in general 
infinitely many energy levels, so it is natural to use this possibility. 
We are studying a quantum computation based on multi--level systems of atoms 
(a qudit theory), \cite{KF2}--\cite{KuF}. Therefore we would like to extend 
the Tavis--Cummings model based on two--levels to a model based on 
multi--levels. This is a very challenging task !

\par \vspace{10mm}
\begin{center}
 \begin{Large}
   {\bf Appendix}
 \end{Large}
\end{center}
\vspace{5mm}
In this appendix we show an another approach to obtain the result in the 
two atoms case which may be useful in the three atoms one. Our method is 
to reduce the 
$4\times 4$--matrix $A$ in (\ref{eq:A-two}) to a $3\times 3$--matrix $B$ 
in (\ref{eq:B-two}) to make our calculation easier.
For that aim we prepare the following two matrices 
\[
T=
\left(
  \begin{array}{cccc}
    1 &                    &                     &    \\
      & \frac{1}{\sqrt{2}} & -\frac{1}{\sqrt{2}} &    \\
      & \frac{1}{\sqrt{2}} &  \frac{1}{\sqrt{2}} &    \\
      &                    &                     & 1
  \end{array}
\right),\quad 
S=
\left(
  \begin{array}{cccc}
      & 1 &   &    \\
    1 &   &   &    \\
      &   & 1 &    \\
      &   &   & 1
  \end{array}
\right).
\]
Then it is easy to see 
\[
S\left(TAT^{\dagger}\right)S^{\dagger}=
\left(ST\right)A\left(ST\right)^{\dagger}=
\left(
  \begin{array}{cccc}
    0  & 0                   & 0                   & 0          \\
    0  & 0                   & \sqrt{2}a           & 0          \\
    0  & \sqrt{2}a^{\dagger} & 0                   & \sqrt{2}a  \\
    0  & 0                   & \sqrt{2}a^{\dagger} & 0
  \end{array}
\right)\equiv 
\left(
  \begin{array}{cc}
     0 &   \\
       & B 
  \end{array}
\right)
\]
where 
\begin{equation}
\label{eq:B-two}
B\equiv 
\left(
  \begin{array}{ccc}
    0                   & \sqrt{2}a           & 0          \\
    \sqrt{2}a^{\dagger} & 0                   & \sqrt{2}a  \\
    0                   & \sqrt{2}a^{\dagger} & 0
  \end{array}
\right)
=
J_{+}\otimes a + J_{-}\otimes a^{\dagger}. 
\end{equation}
$\left\{J_{+},J_{-}\right\}$ are just generators of (spin one) irreducible 
representation of (\ref{eq:sigmas}). Therefore to calculate 
$\mbox{e}^{-itgA}$ we have only to do $\mbox{e}^{-itgB}$. 
The method is almost similar. Namely, 
\[
B^{2}=2
\left(
  \begin{array}{ccc}
    N+1               & 0    & a^{2}   \\
    0                 & 2N+1 & 0       \\
    {a^{\dagger}}^{2} &      & N
  \end{array}
\right),\ 
B^{3}=
\left(
  \begin{array}{ccc}
    2(2N+3) &         &          \\
            & 2(2N+1) &          \\
            &         & 2(2N-1)
  \end{array}
\right)B\equiv DB,
\]
so we obtain 
\begin{equation}
\label{eq:solution-two-more(reduced)}
\mbox{e}^{-itgB}=
{\bf 1}+
\frac{1}{D}\left\{-{\bf 1}+\mbox{cos}\left(tg\sqrt{D}\right)\right\}B^{2}
-i\frac{1}{\sqrt{D}}\mbox{sin}\left(tg\sqrt{D}\right)B
=
\left(
  \begin{array}{ccc}
    b_{11} & b_{12} & b_{13} \\
    b_{21} & b_{22} & b_{23} \\
    b_{31} & b_{32} & b_{33}
  \end{array}
\right)
\end{equation}
where 
\begin{eqnarray}
b_{11}&=&\frac{N+2+(N+1)\mbox{cos}\left(tg\sqrt{2(2N+3)}\right)}{2N+3},\quad 
b_{12}=-i\frac{\mbox{sin}\left(tg\sqrt{2(2N+3)}\right)}{\sqrt{2N+3}}a,
\nonumber \\
b_{13}&=&\frac{-1+\mbox{cos}\left(tg\sqrt{2(2N+3)}\right)}{2N+3}a^{2},\quad 
b_{21}=
-i\frac{\mbox{sin}\left(tg\sqrt{2(2N+1)}\right)}{\sqrt{2N+1}}a^{\dagger},
\nonumber \\
b_{22}&=&\mbox{cos}\left(tg\sqrt{2(2N+1)}\right),\quad 
b_{23}=-i\frac{\mbox{sin}\left(tg\sqrt{2(2N+1)}\right)}{\sqrt{2N+1}}a, 
\nonumber \\
b_{31}&=&
\frac{-1+\mbox{cos}\left(tg\sqrt{2(2N-1)}\right)}{2N-1}{(a^{\dagger})^2},
\quad 
b_{32}=-i\frac{\mbox{sin}\left(tg\sqrt{2(2N-1)}\right)}{\sqrt{2N-1}}
a^{\dagger},
\nonumber \\
b_{33}&=&\frac{N-1+N\mbox{cos}\left(tg\sqrt{2(2N-1)}\right)}{2N-1}. \nonumber
\end{eqnarray}
We leave the remainder to the readers.


\vspace{5mm}
{\bf Note added in the text}

\par
After submitting this paper Pablo P. Munhoz kindky informed us of the paper 

\par \noindent
M.S. Kim, J. Lee, D.Ahn, and P.L. Knight : Entanglement induced by a 
single-mode heat environment, Phys. Rev. A 65, 040101 (2002), quant-ph/0109052 

\par \noindent
and suggested seeing the formula (4). We have checked the agreement 
between the formula (4) and our result, so our result is not new. 
However, how to derive the formula is not written in the paper and 
our method seems much simpler (or easy to understand), so our paper is still 
valuable. 

We wish to thank Pablo P. Munhoz for his helpful comment.

\end{document}